\documentclass[aastex,superscriptaddress,showpacs,letterpaper,showkeys,preprintnumbers,altaffilletter,amssymb,amsmath,amsfonts,aps]{emulateapj}

\usepackage{amsmath}
\usepackage{color}
                                                                                                                                                                                     
\usepackage{natbib}
\shorttitle{Prospects for joint GW and GRB observations}
\shortauthors{Clark, et. al.}

\begin{document}
\title{Prospects for joint gravitational wave and short gamma-ray burst observations}                                                                                                                                                                        
\author{J. Clark\altaffilmark{1,2}, H. Evans\altaffilmark{1}, S. Fairhurst\altaffilmark{1}, 
I. W. Harry\altaffilmark{1,3} , E. Macdonald\altaffilmark{1}, D. Macleod\altaffilmark{1,4}, 
P. J. Sutton\altaffilmark{1}, A. R. Williamson\altaffilmark{1} }

\altaffiltext{1}{School of Physics and Astronomy, Cardiff University, Cardiff, UK} 
\altaffiltext{2}{University of Massachusetts Amherst, Amherst, MA 01003, USA }
\altaffiltext{3}{Department of Physics, Syracuse University, Syracuse NY} 
\altaffiltext{4}{Louisiana State University, Baton Rouge, LA 70803, USA }

\begin{abstract}

We present a detailed evaluation of the expected rate of joint gravitational-wave and short gamma-ray burst 
(GRB) observations over the coming years.  We begin by evaluating the 
improvement in distance sensitivity of the
gravitational wave search that arises from using the GRB observation to restrict the time and sky location of the source.
We argue that this gives a 25\% increase in sensitivity when compared to an 
all-sky, all-time search, corresponding to 
more than doubling the number of detectable gravitational wave signals associated with GRBs.  Using this, we 
present the expected rate of joint observations with the advanced LIGO and Virgo instruments, taking into account the 
expected evolution of the gravitational wave detector network.  We show that in 
the early advanced gravitational wave 
detector observing runs, from 2015-2017, there is only a small chance of a 
joint observation.  However, as the detectors 
approach their design sensitivities, there is a good chance of joint observations provided wide field GRB satellites,
such as Fermi and the Interplanetary Network, continue operation.  The rate 
will also
depend critically upon the nature of the progenitor, with neutron star--black hole systems
observable to greater distances than double neutron star systems.  The relative rate of binary mergers and GRBs 
will depend upon the jet opening angle of GRBs.  Consequently, joint observations, as well as accurate measurement 
of both the GRB rate and binary merger rates, will allow for an improved estimation of the opening angle of GRBs.

\end{abstract}

\section{Introduction}\label{sec:intro}

Two distinct classes of gamma ray bursts were first proposed by 
\cite{Kouveliotou:1993yx}, who argued for distinct populations of bursts with different durations and spectral
hardness.  The separation between short and long bursts is typically taken to be a duration of two seconds.
Long GRBs have subsequently been associated with the death of massive stars, based on the localization
of afterglows to the star forming regions of galaxies \citep{Bloom:1998di} and the association of long 
GRBs with supernovae \citep{Hjorth:2003jt}. Mergers of binary neutron star (BNS) or neutron star--black hole (NSBH)  
systems are a strong candidate for the progenitor of short GRBs \citep{Eichler:1989ve}. 
There is a large amount of evidence in support of the binary merger progenitor hypothesis \citep{Berger:2013jza}, although
nothing definitive.

Binary mergers are also strong emitters
of gravitational waves (GW), at frequencies which the LIGO and Virgo detectors have good sensitivity \citep{thorne.k:1987}.
Consequently, it 
makes sense to search for gravitational-wave signals originating at a time and sky position consistent 
with the observed GRB signal.
This significantly reduces the size of the gravitational wave parameter space by restricting the time, 
sky location and component masses of the binary \citep{Williamson:2014wma}.  An observed signal would allow for the
unambiguous identification of a binary merger origin of a short GRB, providing the strongest possible
backing for the favoured progenitor model.
To date, numerous searches for gravitational waves associated with short GRBs \citep{2012ApJ...760...12A, Aasi:2014iia}
have been performed with the data from the initial LIGO and Virgo detectors \citep{Abbott:2007kv, Acernese:2007zze}. 
Overall, gravitational wave searches have been performed around the time of 80 short GRBs \citep{Aasi:2014iia},
with no evidence of a signal.  Given the sensitivity of the detectors (tens of Mpc for BNS and NSBH mergers) and the
typical measured redshifts of short GRBs (median z=0.4), this was to be expected.
There were, however, two short GRBs, GRB\,051103 
\citep{2010MNRAS.403..342H} and GRB070201 \citep{2007GCN..6103....1H}, whose sky locations overlapped nearby galaxies.  
The non-detection of gravitational waves associated with these GRBs \citep{Abadie:2012bz, Abbott:2007rh} provided weight to the argument that these events were extra-galactic giant magnetar flares.

The second generation of gravitational wave interferometric detectors, Advanced LIGO \citep{TheLIGOScientific:2014jea}
and Advanced Virgo \citep{Acernese:2015gua}, are under construction and expected to be 
operational in 2015 and 2016 respectively, approaching 
design sensitivities over 3--5 years \citep{Aasi:2013wya}. 
They will provide a factor of ten increase in sensitivity over a broad range of frequencies, and will therefore be
sensitive to binary mergers within a few hundred Mpc for BNS and up to 1 Gpc for 
NSBH\citep{Abadie:2010cf}, comparable with
the distances to the closest short GRBs.  
The observation of a gravitational wave signal in coincidence with a GRB is therefore a realistic prospect.
Such an observation would firmly establish binary mergers
as the progenitors of short GRBs and also allow us to distinguish a BNS or NSBH progenitor in many cases.  
This paper discusses in detail the prospects of such a joint 
observation as the advanced detector network evolves towards its design sensitivity.

In order to accurately evaluate the detection prospects, we must evaluate the expected rates of short GRBs in the local
universe and the sensitivity and sky coverage of both the GRB and gravitational wave detectors.  Numerous studies have provided estimates of the rates of binary mergers and GRBs within the range
of the advanced gravitational wave detectors (see, for example \cite{Abadie:2010cf} or \cite{Wanderman:2014eza}).  
Using these, it is possible to estimate the
fraction of GRBs which will produce an observable gravitational wave signal.  In addition to GRB rates and
detector sensitivities,  this
will also depend upon the masses of the progenitors (and in particular whether they are BNS or NSBH systems) as
well as the beaming angle.  We obtain an estimate of the rate of joint observations, and show that there is a greatly increased chance of observing a GW around the time of a GRB, in comparison to an arbitrary stretch of data.  This increased detection 
probability allows for a reduction of the detection threshold, to maintain a fixed false detection probability.  A timeline for the expected evolution in sensitivity of the advanced LIGO (including LIGO India) and Virgo detectors is given in \cite{Aasi:2013wya}.  We use this to calculate the expected rate of joint GW-GRB observations as the detectors evolve towards their full sensitivities.

The measure of a GRB redshift provides an additional piece of information, above and beyond the observed
time and sky location, that can be used to restrict the parameters of the gravitational wave search.  This will
be valuable as the vast majority of observed GRBs are expected to be outside the range of the gravitational 
wave detectors.  By incorporating this information into the gravitational wave search, it should be possible
to increase the sensitivity of the search to GRBs with measured redshift.

The electromagnetic emission from short GRBs is believed to be be beamed.  The jet beaming angle can be
measured by observing a jet break in the electromagnetic emission \citep{Sari:1999mr}, and several opening
angles have been measured to be less than $10^{\circ}$, with other short GRBs having lower limits up to $20^{\circ}$
\citep{Berger:2013jza}.  It will be difficult to constrain the opening angle based on the observation of a gravitational 
wave signal, due to the low signal to noise ratio of the
observations, and an inability to independently measure distance and binary inclination. 
However, the beaming angle will also affect the relative rates of observed binary 
mergers and short GRBs: more tightly beamed emission will lead to a lower fraction of binary mergers with 
observable gamma-ray emission.  Consequently, the accurate measurement of both the local GRB rate
and the binary merger rate will allow us to infer the (average) beaming angle.  Even in the absence of a 
gravitational wave detection, we can place lower bounds on the short GRB jet opening angle.  We also present the 
expected bounds from early advanced detector runs.

This paper is laid out as follows: in section \ref{sec:short_grb} we discuss GRB observations, briefly 
reviewing the evidence for the binary merger model of short GRBs and detailing the GRB model we use to obtain
our results.  In section \ref{sec:gw_searches}, we discuss targeted gravitational wave searches, providing
a brief discussion of previous searches and an evaluation of the sensitivity improvement afforded by a GRB observation.
We present the prospects for joint observations in section \ref{sec:joint_obs} 
and in section \ref{sec:payoff} we discuss the
benefits of these observations, focusing on redshift measurements and constraining the opening angle.  We end, in
section \ref{sec:discussion}, with a discussion.

\section{Observations of short GRBs}
\label{sec:short_grb}

\subsection{GRB satellites and observations}
\label{sec:grb_obs}

The first GRBs were observed by the Vela satellites \citep{Klebesadel:1973iq}, although
 it was not until the 
Burst and Transient Source Explorer (BATSE) instrument on the Compton Gamma Ray
Observatory, that they were shown to be of cosmological origin, and classified into two families.

GRB observations were revolutionised by the Swift satellite which, as well as a large area 
Burst Alert Telescope (BAT), also carries sensitive X-ray and UV-optical telescopes which can be slewed
rapidly to observe the burst afterglow \citep{2004ApJ...611.1005G}.  
Swift has been operational since 2004, has 
detected over 800 GRBs to date, including 75 short bursts,
and has a field of view of approximately 2\,sr.  The Swift BAT observes around 10 short 
GRBs per year%
\footnote{http://swift.gsfc.nasa.gov/}.  
It typically gives localizations with arc minute accuracy, or better if the burst is followed up with the onboard 
X-ray telescope.  Swift's ability to localise sources rapidly and accurately has enabled the follow-up observation 
of numerous short burst afterglows, measurements of redshifts and identification of galaxy hosts.  The Swift satellite 
is expected to continue operations until at least 2020.

The Fermi satellite was launched in 2008 and carries on board the Gamma-ray Burst Monitor (GBM) 
and the Large Area Telescope (LAT)%
\footnote{http://fermi.gsfc.nasa.gov/}.  It is GBM which provides the broadest sky coverage and is 
essentially an ``all--sky'' telescope with a  field of view of 9.5\,sr 
\citep{Meegan:2009qu}
Fermi GBM typically observes around 45 short bursts per year, of which only a small fraction are seen in LAT.
The Fermi localisation is typically accurate to tens or hundreds of square degrees \citep{0067-0049-211-1-13},
making optical followup of these events challenging. To date, no afterglow from a short GRB observed only by Fermi 
has been observed, and consequently the redshifts of these bursts are not known. 
Fermi is currently operational, with its 10 year funding cycle ending in 2018, though it may continue 
operations further. 

The Space-based multi-band astronomical Variable Objects Monitor (SVOM) satellite%
\footnote{http://www.svom.fr/}
is a recently approved Chinese-French mission that is scheduled for launch in 2021 
\citep{Basa:2008cw}.  SVOM will have a similar sky
coverage to Swift, and will also carry X-ray, optical and UV telescopes that can be rapidly and automatically slewed
to observe afterglows. 

Finally, the set of instruments which make up the InterPlanetary Network (IPN)%
\footnote{http://heasarc.gsfc.nasa.gov/W3Browse/all/ipngrb.html}
, are not dedicated GRB satellites, 
but 
instead have GRB monitors on board \citep{Hurley:2002wv}.  The majority of satellites in the network are unable 
to localize the bursts individually but it is possible to localize bursts observed in numerous satellites using triangulation.
The sizes and shapes of these
error regions vary greatly, depending upon the number of satellites and their locations (more distant satellites 
greatly
improve localization).  The IPN provides essentially all sky coverage for GRBs, although, given the sensitivity of 
the
detectors, the GRBs observed tend to be brighter.

There are several key pieces of observational evidence to support the binary merger model for short GRBs.  
The afterglows of several short GRBs have been
observed in both the X-ray \citep{Gehrels:2005qk} and optical \citep{Hjorth:2005kf} and consequently
localized to galaxies.  
Despite several low redshift observations, there has been no observation of a supernova associated with
any of these events.  Additionally, the host galaxies of short GRBs are much more varied than long, with a
large fraction of late type galaxies which are not observed as hosts of long GRBs.  In the binary merger
model, the delay time between formation and merger of the binary can take a wide range of values 
\citep{Dominik:2012kk}, explaining the range of galaxy types observed as hosts of short GRBs.  There is also
evidence of ``hostless'' short GRBs that have been ejected from the galaxy \citep{Fong:2013iia} which
arise naturally from supernova kicks imparting a velocity to the binary, coupled with the 
long delay time to merger.  The strongest
evidence for the merger model is the observation of a kilonova associated with GRB130603B \citep{Tanvir:2013pia, 
Berger:2013wna}.  Taken together, these 
observations provide good evidence for the binary merger model for at least a subset of short GRBs.  
We note, however, that some fraction may still be mis-classified long GRBs \citep{Bromberg:2012gp} or
soft gamma repeaters in nearby galaxies \citep{Abbott:2007rh, 2010MNRAS.403..342H}. 

\subsection{The rate of short GRBs}

There have been numerous
efforts recently to estimate the rate of short GRBs, based primarily on redshift measurements of 
GRBs observed by Swift \citep{2012MNRAS.425.2668C, Wanderman:2014eza, 2014MNRAS.437..649S}.%
\footnote{A nice summary of recent rate estimates is provided in Table 4 of \citep{Wanderman:2014eza}}
Here, we follow \cite{Wanderman:2014eza}, who use the observed GRB populations 
(and measured redshifts in Swift) in order to derive a luminosity function for GRBs as well 
as a local rate density.

The energy spectra of short GRBs is modelled, following \cite{1993ApJ...413..281B}, as a power law decay with exponential
cutoff at low energy and a steeper power law at higher frequencies.  The parameters used in the Band function are 
$\alpha_{\mathrm{BAND}} = -0.5$, $\beta_{\mathrm{BAND}} = -2.25$ and $E_{\mathrm{peak}}= 800 \mathrm{keV}$.
For a GRB at a given distance/redshift, the peak photon count in a detector can be related to the peak luminosity in
a straightforward way \citep{Wanderman:2014eza, Regimbau:2015}.  The detection threshold is taken to be $2.5$ photons per second 
in the $15-150$ keV band for Swift and $2.37$ photons per second in the $50-300$ keV band for Fermi.

The short GRB luminosity function is taken to be a broken power law, with a logarithmic distribution
\begin{equation}
\phi_{o}(L) = \left\{ 
\begin{array}{ll}
 \left(\frac{L}{L_{\star}}\right)^{-\alpha_{L}} & L < L_{\star} \\
 \left(\frac{L}{L_{\star}}\right)^{-\beta_{L}} & L > L_{\star} 
\end{array}
 \right.
\end{equation}
where $L$ is the peak luminosity (in the source frame) between $1$ keV and $10$ MeV, and $\alpha_{L}$ and $\beta_{L}$ give the power law decay below and above the break at $L_{\star}$.%
\footnote{Other papers use a smaller energy band when defining the luminosity, and this has an impact on the value of $L_{\star}$,
although not on the slopes of the power law components.}
The other important parameter is the minimum GRB luminosity, which determines the lower cutoff of the luminosity distribution.
This is poorly constrained as only nearby low luminosity GRBs would be observable.  The minimum luminosity is
taken to be $L_{min} = 5 \times 10^{49}$ erg/s.

The parameters $\alpha_{L}$, $\beta_{L}$, $L_{\star}$ are fitted jointly with the short GRB rate.  Best fit values are 
$\alpha_{L}=1$, $\beta_{L}=2$ and $L_{\star} = 2\times 10^{52}$erg/s, with a local GRB rate of $4.1 \mathrm{Gpc}^{-3} y^{-1}$.
The GRB rate evolves with redshift, peaking at $z \approx 1$.

Other works \citep{2012MNRAS.425.2668C, 2014MNRAS.437..649S} take a similar approach to estimating the rate of short GRBs,
although the assumptions they make vary.  Consequently there is some variation in the rate estimates, but they typically lie in the range 
$1-10 \times 10^{-9}\, \mathrm{Mpc}^{-3} \mathrm{y}^{-1}$ with a median rate around $3 \times 10^{-9}\, 
\mathrm{Mpc}^{-3} \mathrm{y}^{-1}$.  These rates are somewhat lower than earlier estimates based on a
smaller sample of GRBs \citep{Guetta:2005bb, Nakar:2007yr}.  
For the remainder of this work, we make use of the Band function and luminosity distribution parameters of \cite{Wanderman:2014eza}, 
but allow for a constant rate per comoving volume between $1$ and $10$ per Gpc${}^{3}$ yr.  We do not include any variation of
GRB rate with redshift as we found it had little impact on the overall results, due to the limited range of the gravitational wave detectors.

Given the evidence for a binary merger progenitor for short GRBs, it is interesting to compare
the observed and predicted rates of short GRBs and binary mergers.  
To do so, we must take into account
the beaming of the GRB jet.  The evidence for beaming in short GRBs comes primarily from the 
observation of jet breaks, at which time the material in the jet starts to spread out, leading to 
a break in the light curve.  The observation of such a break can be used to infer the
jet's opening angle \citep{Sari:1999mr}.  The observation of a jet break in a number of 
short GRB afterglows (see for example \cite{Fong:2013lba, Panaitescu:2005er, 
Guelbenzu:2012id}) has been used to infer opening angles between $3^{\circ}$ and $8^{\circ}$.  
In others, the lack of an observed break has been used to set a lower limit on the beaming angle.  
In many cases this leads to a limit of only a few degrees, however, GRB 050724 had no observed break 
after 22 days, leading to an inferred opening angle of at least $20^{\circ}$.  See \cite{Berger:2013jza} for a
summary of observations to date.

The rate of observed short GRBs can be related to the all sky rate of binary mergers through
\begin{equation}\label{eq:rates}
  R_{\mathrm{GRB}} = f_{\gamma} (1 - \cos{\theta_{j}}) R_{\mathrm{merger}} \, 
\end{equation}  
where $\theta_{j}$ is the average jet opening angle of the gamma ray emission, and the factor $f_{\gamma}$ 
encodes the fraction of binary mergers which produce a gamma ray burst.  
The rate of neutron star binary (BNS) mergers, inferred from binary pulsar observations and
population synthesis modelling, is taken to lie between $1 \times 10^{-5}$ and  
$1 \times 10^{-8} \mathrm{Mpc}^{-3} \mathrm{y}^{-1}$ (see \cite{Abadie:2010cf} and references therein). To date, no 
neutron star--black hole (NSBH) systems have been observed as binary pulsars, but the
rate can still be predicted through population synthesis, constrained by the observations of
double neutron star binaries, to be $10^{-6} - 6 \times 10^{-10} \mathrm{Mpc}^{-3} \mathrm{y}^{-1}$.

\begin{figure}[tb]
\begin{center}
\includegraphics[width=\linewidth]{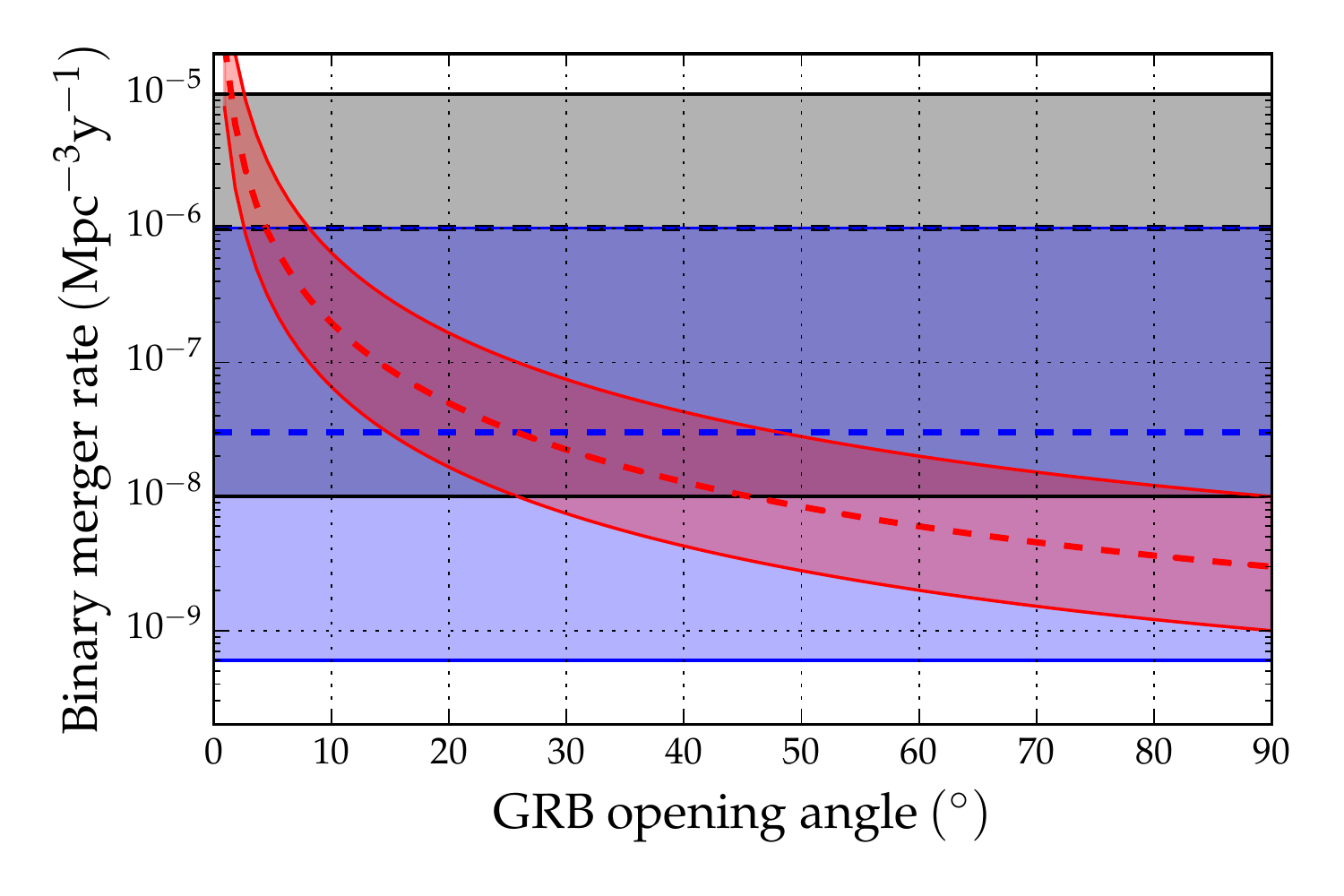}
\caption{\label{fig:rate_prediction}
The rate of binary mergers in the local universe.  The figure shows the predicted rates of binary neutron star (upper, grey band)
andneutron star black hole (lower, blue band) mergers, taken from \cite{Abadie:2010cf}.  The shaded regions mark the range of 
reasonable values, while the dashed lines show the best estimate of the rate.  We can also infer 
the rate of GRB progenitors, given an opening angle, as plotted in red.}
\end{center}
\end{figure}

In Fig. \ref{fig:rate_prediction}, we compare the observed and predicted rates for short GRBs 
to those for BNS and NSBH mergers.  As has been observed elsewhere \citep{Guetta:2005bb}, 
there is a remarkable concordance
between the GRB and BNS rates, with the observed beaming angles compatible with
the best guess BNS rate, with a lower rate of BNS mergers requiring larger GRB opening
angles.%
\footnote{For this discussion, we have implicitly been assuming that \textit{all} BNS mergers, produce
GRBs, i.e. $f_{\gamma} = 1$ in (\ref{eq:rates}).  There are, however, arguments that only a subset of
BNS mergers will produce GRBs, for example \cite{Giacomazzo:2012dg}.}%
 For NSBH, the rates are not in such good agreement.  A $5^{\circ}$ jet angle
requires an NSBH rate right at the top of the predicted range; alternatively a lower NSBH rate is consistent
with a wider opening angle than has been inferred from observations.  Furthermore, it is likely that a 
a reasonable fraction of NSBH mergers will not produce any electromagnetic emission as the NS
will be swallowed whole, leaving no material from which to form an accretion disk and, hence, GRB jet 
\citep{2012PhRvD..86l4007F, Pannarale:2014rea}.  Indeed, in \cite{Stone:2012tr}, the fraction, $f_\gamma$, of NSBH mergers that produce short GRBs is argued to be between $0.1$ and $0.3$, depending upon black hole mass and spin distributions.  Thus, based on rate estimates, it seems unlikely that 
NSBH mergers can account for \emph{all} observed short GRBs.  

\section{Gravitational Wave searches for GRB progenitors}
\label{sec:gw_searches}

\subsection{Implementation of a targeted search}

The details of the targeted search for gravitational waves associated with 
short GRBs are 
presented in \cite{Harry:2010fr, Williamson:2014wma}.  Given the time and sky location of the GRB,
the search is restricted to a six second span of data around the GRB, allowing for a 
merger time 5 seconds before and 1 second after the time of the GRB. This 
allows for most
realistic delays between the merger and GRB signal.  The data from the available 
gravitational wave detectors is combined coherently, by appropriately time shifting and weighting
the data from each detector to account for the known sky location of the source, to produce
data streams sensitive to the two gravitational wave polarisations.  The data is then searched
through matched filtering using a template bank of binary merger waveforms \citep{lrr-2014-2, Owen:1998dk}.  
GRBs observed by Swift are localised to well under a square degree which, for the gravitational wave search
means they are effectively localised to a single sky location. 
For  GRBs observed by Fermi GBM or the InterPlanetary Network, the sky location may be
poorly constrained by satellite observations, with uncertainties of tens to hundreds of square degrees.  
In this case, a grid of sky points is searched \citep{Predoi:2011aa, Williamson:2014wma}.  Additional
signal consistency tests \citep{Allen:2005kaa, Harry:2010fr}, designed to eliminate glitches which
have a different signal morphology than binary merger signals, are used to eliminate spurious events 
due to non-Gaussian noise. The background is evaluated from data surrounding the six second stretch used 
in the search and simulated signals are added to the data to evaluate the search 
sensitivity.

\subsection{The benefit of a targeted search}
\label{sec:sens_improv}

The observation of a short GRB provides a good estimate of the merger time, 
sky location and (possibly) distance of a potential binary merger signal.  This 
significantly reduces the parameter space of a follow-up gravitational wave
search and consequently allows for a
reduction in the detection threshold 
\citep{Dietz:2012qp, Kelley:2012tc, Chen:2012qh}.  We evaluate in detail
the sensitivity improvement afforded by the GRB observation.  In contrast to most previous
studies, we will make use of the results obtained from searches on real data and
make use of the results of previous analyses \citep{Colaboration:2011np, Williamson:2014wma}.

To investigate the impact of reducing the parameter space for GRB searches, 
we will deliberately avoid the
question of first gravitational wave detection --- where a ``5-sigma'' observation
may well be required \citep{Colaboration:2011np}.  
Instead, we consider a later observation for which we might require a specific
false positive rate: i.e. a limit on the fraction 
gravitational wave observations are spurious.  In that case, the threshold for
announcing a detection is tied to the true signal rate.  Since neither the GRB or BNS rates 
are known with great accuracy, for this discussion we will 
adopt with the ``realistic'' rates of $10^{-6} \mathrm{Mpc}^{-3} \mathrm{y}^{-1}$ for BNS 
mergers and $3 \times 10^{-9} \mathrm{Mpc}^{-3} \mathrm{y}^{-1}$ for short GRBs.

A detailed evaluation of the expected rate of BNS observations is provided in \cite{Aasi:2013wya}.
There, a false rate of one event per century is chosen, corresponding to a signal to noise ratio
of 12 in the advanced detectors.  When the advanced LIGO and Virgo detectors are operating
at design sensitivity, the expected rate of observed BNS mergers is 20 per year.  Thus the threshold 
corresponds to a false positive rate of 1 in 2000.  

To obtain a comparable SNR threshold for the GRB search,  we need
to evaluate both the expected foreground and background around the time of a short GRB. Using
the results of \cite{Williamson:2014wma} we estimate a background rate of 1 in 1,000 for event with an SNR above 8
in the GRB search, with the background decreasing by a factor of 100 for a unit increase in SNR:%
\footnote{The analysis in \cite{Williamson:2014wma} was performed for the initial LIGO and Virgo detectors and,
assuming that GRB emission is beamed and the jet is perpendicular to the plane of the binary, we
obtain a background of 1 in $10^{5}$ above SNR of 8.  However, we must include a trials factor due
to requiring a larger template bank for the advanced detectors \citep{Owen:1998dk} and consequently
we (somewhat conservatively) increase the background by a factor of 100 as was done in \cite{Aasi:2013wya}}
\begin{equation}\label{eq:fap}
P_\mathrm{BG}(\rho > \rho_{\star}) = \left\{
\begin{array}{ll}
 \displaystyle 10^{-(5 + 2[\rho_{\star}-9])} & \rho_{\star} > 6.5 \\
 \displaystyle 1 & \rho_{\star} \le 6.5 \, .
\end{array}
\right.
\end{equation}

Next, we must determine the probability of any given GRB occurring at a low enough redshift that the
gravitational wave signal would be observable by the advanced LIGO and Virgo network. 
The sky and binary orientation averaged sensitivity of the network is $200 \mathrm{Mpc}$.
However, it is natural to assume that the GRB jet is beamed perpendicular to the plane of the binary's orbit 
(see e.g. \cite{2013MNRAS.430.2121P}).  
The gravitational wave signal is also (weakly) beamed in this direction: the  
amplitude for a face on signal is a factor of 1.5 greater than the orientation averaged amplitude.%
\footnote{The sensitivity of a detector to binary mergers is typically quoted in two different ways: either the 
range --- the sky and orientation averaged sensitivity; or the horizon --- the maximal sensitivity, for binaries
which are directly overhead the detector and face on.  The horizon distance is a factor of 2.26 greater than the
range.  Here, we are assuming all sources are face on, but still averaging over sky positions.  It turns out that
the averaging over orientation and sky give the same factor, so performing just one average increases
the sensitivity by $\sqrt{2.26} = 1.51$.}
The gravitational wave beaming is rather weak and the amplitude falls off slowly with opening angle.
Even with opening angles up to $30^{\circ}$ the mean amplitude 
is only reduced by 5\% from the face on case \citep{Dietz:2012qp}.
Thus, the nominal sensitivity for GRB signals in the advanced detector network is 300 Mpc, rather than
200 Mpc for signals of arbitrary orientation.  The sensitive distance scales inversely with the SNR 
threshold, i.e.
\begin{equation}
  D_{\star} = \left(\frac{12}{\rho_{\star}}\right) 300 \mathrm{Mpc} \, .
\end{equation}

There are around 50 short GRBs observed annually (10 by Swift BAT and 45 by Fermi GBM, of which several are
observed by both instruments).  Assuming a local GRB rate of  
$3 \times 10^{-9} \mathrm{Mpc}^{-3} \mathrm{y}^{-1}$, we would expect around one event per year to be detected at a distance
of 500 Mpc or less, taking into account detector sensitivities, sky coverage and live times.  Thus, the chance of any GRB 
occurring within a distance $D_{\star}$ can be approximated as 
\begin{equation}\label{eq:grb_prob}
  P_{\mathrm{GRB}}(D < D_{\star}) \approx \frac{1}{50} \left(\frac{D_{\star}}{500 \mathrm{Mpc}}\right)^{3} \;\;
  D_{\star} \lesssim 500 \mathrm{Mpc} \, .
\end{equation}
We have ignored the impact of detector sensitivity since, assuming the
GRB model in the previous section, the majority of GRBs within this range would be observed by Swift or Fermi if
they were in the field of view.
This is broadly consistent with the observed redshifts from Swift, where the smallest of 30 measurements 
is z=0.12, corresponding to a distance of $550 \mathrm{Mpc}$.  Obviously, this relationship will break down at 
larger distances where cosmological effects, variation of the intrinsic GRB rate and detection efficiencies all
become significant.

In the GRB search, the chance of a noise event giving an SNR above 9.1 is $5 \times 10^{-6}$.
At this SNR, the sky averaged sensitivity to face on BNS mergers is $400$ Mpc so, from equation (\ref{eq:grb_prob}),
there is a 1\%
chance of the gravitational wave signal from a short GRB being observable. 
This gives a false positive rate of 1 in 2000 as desired.  
Therefore, the observation of a GRB allows us to lower the threshold in a gravitational 
wave search by 25\% while maintaining a fixed false positive rate.  We note that neither the 
astrophysical rate of BNS or GRBs nor the noise background of the advanced detectors are known at this time.
Nonetheless, the predicted increase in sensitivity of the GRB search is relatively robust.  
The observed background for the BNS and GRB searches is very 
similar in nature and, in particular, both show the same, rapid rate of falloff at large SNR.  Thus, changes
in the required detection confidence will affect both searches in the same way.

Reducing the detection threshold by 25\% will more than double the number of detectable signals.
In other words, less than half of gravitational wave signals associated with GRBs will be detected based
on the gravitational wave signal alone --- it is only with a joint search that makes use of the GRB 
observation that these additional signals will be seen.

It is instructive to ask why the detection threshold can be lowered by 25\% for the GRB search.
The answer is twofold.  First, the expected rate of signals is significantly higher in the data around
the time of a GRB.  In equation (\ref{eq:grb_prob}), we gave the probability of there being an observable signal
in the 6 seconds of data around the time of a GRB, as a function of the sensitive distance.
Within the nominal GRB range of 300 Mpc (at SNR 12), there is a 1 in 250 chance of observing a signal
associated to the GRB.  Meanwhile, for an arbitrary six seconds of data, assuming a BNS rate of
 $10^{-6} \mathrm{Mpc}^{-3} \mathrm{y}^{-1}$, there is a 1 in 150,000 chance of observing a signal associated to a
 BNS merger.  Thus, assuming that BNS are GRB progenitors, it is around a thousand times more likely 
that we observe a signal within the 6 seconds around a GRB than in an arbitrary six
seconds of data.  In addition, the GRB background is further reduced because searching a
small time window makes a fully coherent search feasible \citep{Harry:2010fr}, and this 
increases the sensitivity relative to the all sky search \citep{Babak:2012zx}.  
These factors combine to lead to the 25\% reduction
in threshold that can be achieved by the search.

\section{Expected Rate of joint observations}
\label{sec:joint_obs}

The first advanced detector observing runs of the are expected in late 2015, with 
sensitivity reaching the design specification towards the end of the decade.  At 
design sensitivity, the aLIGO range for BNS mergers will be $200 \mathrm{Mpc}$ and for
AdV, $130 \mathrm{Mpc}$.  Around 2022, a third LIGO detector in India is expected to
begin observing with comparable sensitivity \citep{Aasi:2013wya}.
Given the evolution of the advanced detector sensitivities as well as the results of the previous sections, it is
straightforward to evaluate the 
expectations for joint short GRB--gravitational wave observations in the coming years.
We consider three GRB observing scenarios: Swift, Fermi and full sky,
full sensitivity coverage.  While the latter is, of course, somewhat optimistic, it serves to provide
an upper bound on the joint observation rate.  For Swift and Fermi, we use the sky coverage and 
detection thresholds outlined in section \ref{sec:grb_obs} and, in addition, we assume an 80\% detector 
duty cycle for both detectors due to passage through the South Atlantic Anomaly. 

\begin{deluxetable*}{rrrrrrrr}
\tablecolumns{8}
\tablecaption{The expected rate of joint gravitational wave--GRB observations in the upcoming science runs, assuming that the 
progenitor of every short GRB is a BNS merger. 
Sensitivities, run durations and BNS rates taken from \citep{Aasi:2013wya}.\label{tab:nsns_rate}}
\tablehead{
\colhead{Epoch} & \colhead{Run Duration} & \multicolumn{2}{c}{BNS Range (Mpc)} & 
\multicolumn{3}{c}{Number of GW--GRB detections} \\
\cline{3-4} \cline{6-8} \\
\colhead{} &\colhead{} & \colhead{LIGO} & \colhead{Virgo} & 
\colhead{All Sky} & \colhead{Fermi GBM} & \colhead{Swift BAT} }
\startdata
2015        & 3 months & 40 - 80   &     -        & 
$2 \times 10^{-4}$ - 0.02  & $2 \times 10^{-4}$ - 0.02 & $3\times 10^{-5}$ - 0.003 \\
2016--17 & 6 months & 80 - 120 & 20 - 60   & 
0.004 - 0.2        & 0.003 - 0.1     & $3\times 10^{-4}$ - 0.03\\ 
2017--18 & 9 months & 120-170 & 60 - 85   & 
0.02 - 0.8           & 0.01 - 0.5         & $7\times 10^{-4}$ - 0.1 \\ 
2019+      & (per year) & 200       & 65 - 130 & 
0.1 - 2             & 0.07 - 1           & 0.01 - 0.2 \\
2022+       & (per year) & 200      & 130        & 
0.2 - 3           & 0.1 - 2 & 0.02 - 0.3 
\enddata
\end{deluxetable*}

The expected rates of short GRB observations, assuming a BNS progenitor, are given in Table 
\ref{tab:nsns_rate}. 
For each observing run, a range of possible detector sensitivities is quoted, to take into account the 
uncertain nature of commissioning and operating the advanced detectors \citep{Aasi:2013wya}.  
The rate of observed BNS mergers is calculated for a merger rate 
between $10^{-5}$ and $10^{-8} \mathrm{Mpc}^{-3} \mathrm{y}^{-1}$.  The range of predicted rates reflects
the uncertainty in both the detector sensitivities and the rate of sources.  For joint gravitational wave--GRB observations,
we take the short GRB rate to lie in the range  $10^{-8} - 10^{-9} \mathrm{Mpc}^{-3} \mathrm{y}^{-1}$. 
As discussed in section \ref{sec:sens_improv}, we allow for a 25\% decrease in 
detection threshold associated with a dedicated GRB search when compared to an 
all-sky all-time gravitational-wave search. When calculating
the Swift and Fermi rates, we use the GRB luminosity distribution and energy spectra described in section \ref{sec:grb_obs}.
These thresholds, however, have little effect on the rate as the majority of GRBs within the sensitive range of advanced 
LIGO and Virgo will have a peak luminosity sufficient to be observed by Swift BAT and Fermi GBM.

The expected number of joint observations in the early advanced LIGO-Virgo science runs is much less than one.
However, by the 2017--18 observing run, there is a real chance of a joint observation and, with the network operating at 
design sensitivity, an excellent chance of joint gravitational wave--GRB observations during an extended science run.
We note, however, how critical it is to continue monitoring the sky for GRBs:  
it is only with the sky coverage provided by Fermi (and the InterPlanetary Network) that we expect to make
joint observations.

\begin{figure}[tbp]
\begin{center}
\includegraphics[width=\linewidth]{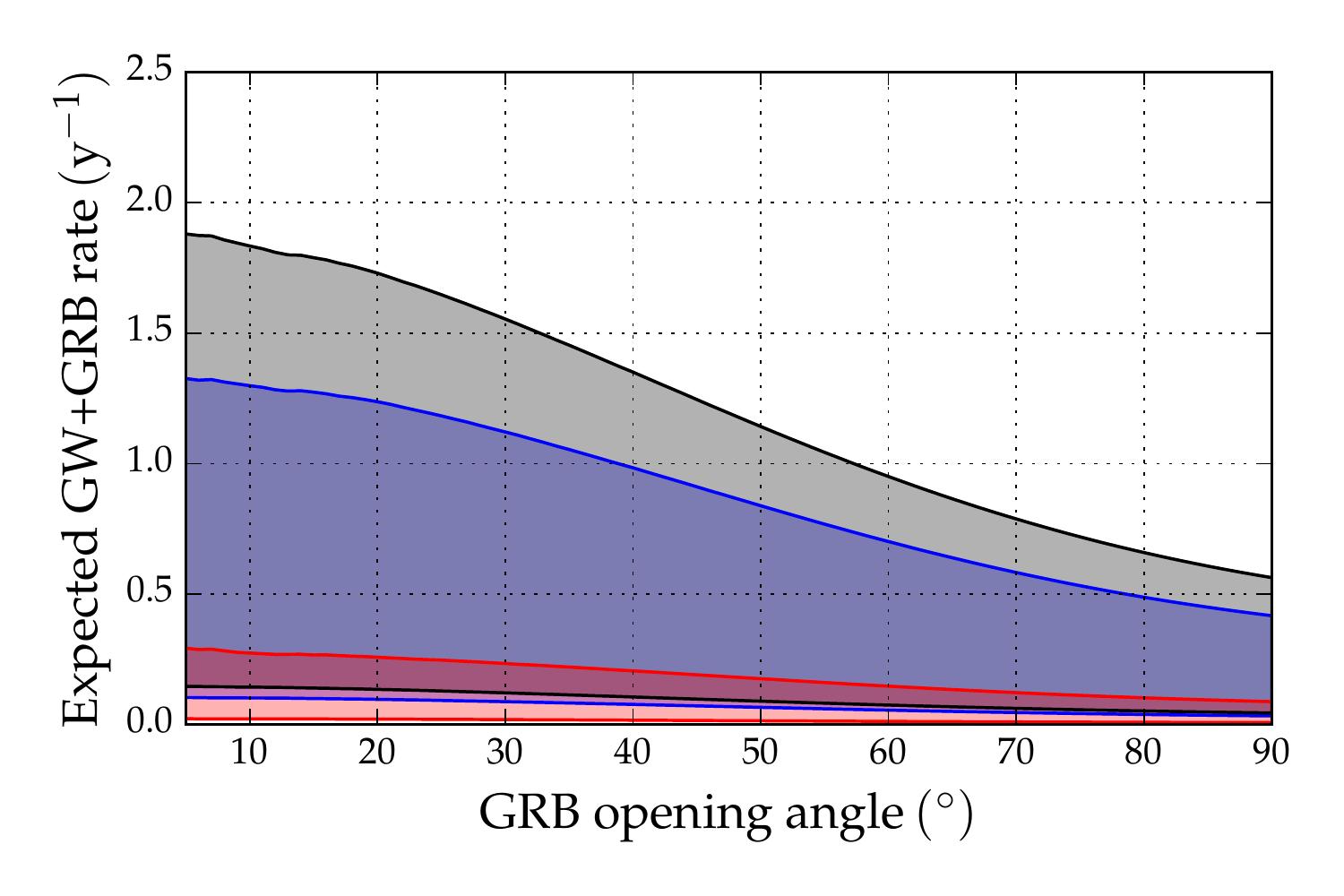}
\caption{\label{fig:bns_grb}
The expected rate of observed gravitational wave--GRB signals when the LIGO and Virgo detectors are operating at
their design sensitivity.  We take the intrinsic short GRB rate to be in the range $1-10 \times 10^{-9} \mathrm{Mpc}^{-3} 
\mathrm{y}^{-1}$ and assume that BNS are the progenitor source of all short 
GRBs.  The grey region shows the range of expected rates with all-sky GRB coverage.  The
observed rate increases with a small opening angle as the systems are close to face on and
thus have the maximum gravitational wave emission.  The blue region shows the expected rate for joint observations
with Fermi GBM and the red region for Swift BAT.  For preferred opening angles (less than $30^{\circ}$) we expect to see at least
one GRB per year in coincidence with Fermi GBM.
}
\end{center}
\end{figure}

\begin{figure}[tbp]
\begin{center}
\includegraphics[width=\linewidth]{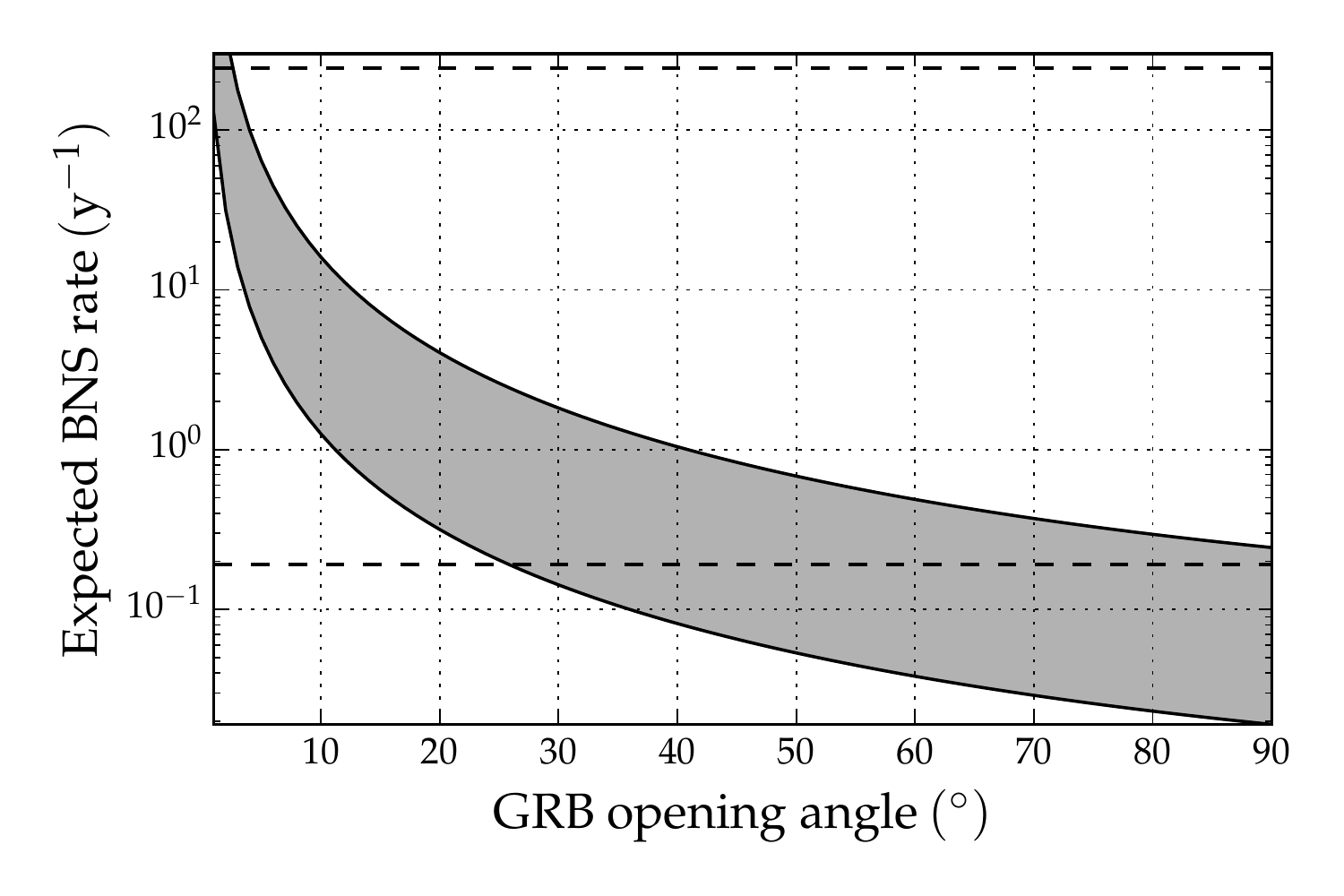}
\caption{\label{fig:bns_all}
The expected rate of observed BNS signals when the LIGO and Virgo detectors are operating at
their design sensitivity.  We take the intrinsic GRB rate to be in the range $1 10 \times 10^{-9} \mathrm{Mpc}^{-3} 
\mathrm{y}^{-1}$.  The rate increases with smaller opening angles as this implies a greater fraction of sources which 
are not observed as GRBs.  The horizontal lines bound the predicted number of observations based upon estimates
of BNS rates.  At the largest opening angles, only the higher GRB rates are consistent with the BNS predictions.}
\end{center}
\end{figure}

Fig.~\ref{fig:bns_grb} shows the expected annual rate of joint observations, as 
a function of GRB opening angle for the 2019+ configuration of Table \ref{tab:nsns_rate}.  The dependence of
the rate on the GRB opening angle is due to the beaming of the gravitational wave signal: the 
amplitude for a face on signal is a factor of 1.5 greater than the orientation averaged signal, giving a 
factor of 3.4 between small opening angles and no beaming.  Fig.~\ref{fig:bns_all} shows the expected 
all sky BNS merger rate, as a function of GRB opening angle
under the assumption that \textit{all} BNS mergers produce gamma-ray emission.
As discussed in \cite{Chen:2012qh}, 
there is a crossover point, where we see more gravitational waves associated with GRBs than in an all
sky, all time search.  This will obviously depend upon the sky coverage and sensitivity of the GRB satellites, but
assuming full sky coverage, this occurs around $40^{\circ}$. If the beaming angle is larger than this,
the GRB search will detect more signals than the all sky all time search, due 
to the ability to lower thresholds around the time of observed GRBs. 
Of course, based on astrophysical measurements
of GRB opening angles, this is unlikely to be the case.

\begin{deluxetable*}{rrrrrrrr}
\tablecolumns{8}
\tablecaption{The expected rate of joint gravitational wave--GRB observations in the upcoming science runs, assuming that the 
progenitor of every short GRB is a NSBH merger. Sensitivities and run durations taken from \citep{Aasi:2013wya}, 
we assume a fiducial NSBH with a neutron star mass of $1.4 M_{\odot}$ and a black hole mass of  $5.0 M_{\odot}$.\label{tab:nsbh_rate}}
\tablehead{
\colhead{Epoch} & \colhead{Run Duration} & \multicolumn{2}{c}{NSBH Range (Mpc)} & 
\multicolumn{3}{c}{Number of GW--GRB detections} \\
\cline{3-4} \cline{6-8} \\
\colhead{} &\colhead{} & \colhead{LIGO} & \colhead{Virgo} & 
\colhead{All Sky} & \colhead{Fermi GBM} & \colhead{Swift BAT} }
\startdata 
2015        & 3 months & 70 - 130 &     -           & 
$3 \times 10^{-4}$ - 0.06 & $2 \times 10^{-4}$ - 0.03 & $4 \times 10^{-5}$ - 0.007 \\
2016--17 & 6 months & 130 - 200 & 30 - 100  & 
0.005 - 0.5     & 0.003 - 0.3     & $7 \times 10^{-4}$ - 0.07 \\ 
2017--18 & 9 months & 200 - 280 & 100 - 140 & 
0.03 - 2         & 0.02 - 1     & 0.004 - 0.3 \\ 
2019+      & (per year) & 330        & 110 - 220 & 
0.2 - 6         & 0.1 - 2        & 0.02 - 0.5  \\
2022+       & (per year) & 330       & 220          & 
0.4 - 10         & 0.2 - 3        &  0.03 - 0.7 
\enddata
\end{deluxetable*}

The expected rates of short GRB observations, assuming a NSBH progenitor are given in 
Table \ref{tab:nsbh_rate}.  For NSBH mergers, the masses and spins of the system have a stronger
effect upon the expected rates of observation.  Higher masses and large, aligned spins result in 
greater gravitational wave emission increasing the distance to which the sources can be
observed.  For simplicity, we take the system to be a 
neutron star of mass $1.4 M_{\odot}$ and a non-spinning black hole of mass $5.0 M_{\odot}$.
Following the same procedure as before, we assume that \textit{all} GRB progenitors are NSBH
binaries and use the GRB model discussed in section \ref{sec:grb_obs} to determine the fraction of GRB
signals that are observed by Swift and Fermi.  This has a significant impact on the rate of observable signals,
particularly in the epochs after 2019.

As we have discussed previously, there is already a tension between the observed GRB rates and predicted
NSBH rates.  Specifically, as is clear from Fig. \ref{fig:rate_prediction}, for 
\textit{all} short GRBs to have
an NSBH origin requires a merger rate at the high end of the predicted range, a relatively
large GRB opening angle, or both.  Additionally, numerical simulations indicate that for a large fraction of NSBH 
mergers, there will not be sufficient matter in the accretion disk to power a GRB, making the rates even
less compatible \citep{2012PhRvD..86l4007F}.  Thus, the assumption that \textit{all} GRBs are due to NSBH 
mergers seems difficult to
accommodate, meaning that the highest rates in Table \ref{tab:nsbh_rate} are 
not realistic.  Nonetheless,
even if 15\% of GRBs have NSBH progenitors, this would double the expected rate of joint observations.
Alternatively, the absence of a joint gravitational wave--GRB observation could be used to limit the fraction of short 
GRBs which have a NSBH progenitor.

To end this section, we compare our results with other recently published works.  \cite{Wanderman:2014eza} calculate the rate of 
joint GRB--gravitational wave detections by simply assuming a 300 Mpc range for the advanced LIGO-Virgo network.  They
obtain a rate of joint Fermi (Swift) observations of $0.4 \pm 0.2$ ($0.06 \pm 0.03$) assuming a minimum peak luminosity of 
$5\times 10^{49}$ erg/s.  This is entirely consistent with the rates for BNS in the 2019+ epoch given in Table \ref{tab:nsns_rate}.  
The fact that they have neglected the directional sensitivity of the gravitational wave
network has little impact as essentially all GRBs within the advanced LIGO-Virgo range will be observable by Swift and Fermi.
By varying the luminosity threshold, they obtain rates that span the same range as ours.  For NSBH systems, they assume a
1 Gpc range for the advanced gravitational wave detectors, compared to our range of 660 Mpc, and consequently obtain a significantly
higher rate ($5 \pm 2$ for Fermi and $0.7 \pm 0.3$ for Swift). \cite{Regimbau:2015} have also calculated joint detection rates of 
gravitational wave--GRB signals.  They predict rates of joint observations with Swift of $0.01-0.5$ per year for BNS and $0.004-0.16$
per year for NSBH.  The rates are broadly comparable to those presented here, although the range goes somewhat higher for 
BNS and lower for NSBH.  These differences arise due to different choices of parameters in the Band function, GRB luminosity
distribution, and detector thresholds.  Additionally, the authors choose a fixed BNS (NSBH) rate of $6 \times 10^{-8} 
(3 \times 10^{-9}) \mathrm{Mpc}^{-3} \mathrm{y}^{-1}$ and a range of opening angles between $5^{\circ}$ and $30^{\circ}$.  With these
rates, NSBH signals could only account for a fraction of GRBs.  This explains why their numbers are lower than the ones in Table 
\ref{tab:nsbh_rate} where we have assumed that \textit{all} GRBs have NSBH progenitors.

\section{Benefits of joint observations}
\label{sec:payoff}

Numerous previous papers have discussed the benefits of joint gravitational wave--GRB observations, including: 
the potential to  confirm (or rule out) the binary merger progenitor model \citep{Eichler:1989ve}; 
measuring the time-delay between the binary merber and the GRB signals to understand jet breakout;
the ability to probe GRB jet opening angles \citep{Dietz:2010eh, Chen:2012qh};
the independent measurement of distance and redshift used as a probe of cosmology 
\citep{Schutz:1986, Nissanke:2009kt}.  We will not discuss all of these in detail, but will
focus on two issues.  First, we discuss how the measurement of a GRB redshift may actually
assist in the detection of a gravitational wave counterpart.  Then, we discuss prospects for measuring or
constraining opening angles.
 
\subsection{Detecting a GRB with measured redshift}

\begin{figure}[tb]
\begin{center}
\includegraphics[width=\linewidth]{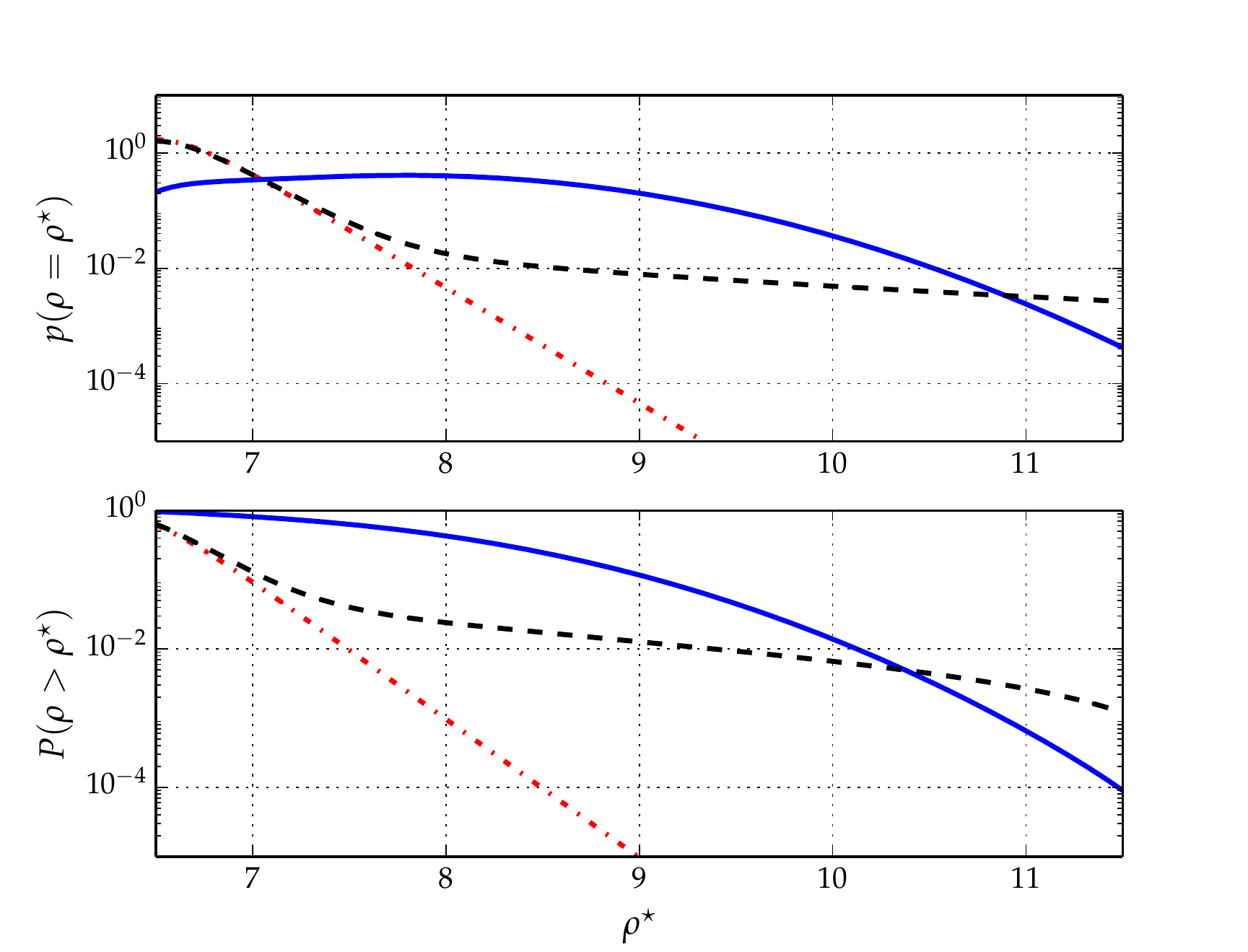}
\caption{\label{fig:fap}
The probability of obtaining an event of a given SNR for: noise only (the red, dot-dashed line); a GRB progenitor at an 
unknown distance (the black, dashed line) and a known distance (the blue, solid line).  In this example, we have used the 
parameters from GRB080905A with a distance of 550 Mpc which gives a signal SNR of 7.7. The top plot shows the 
probability distribution function, while the bottom plot gives the cumulative probability of observing an event as loud 
or louder.}
\end{center}
\end{figure}

\begin{figure}[tbp]
\begin{center}
\includegraphics[width=\linewidth]{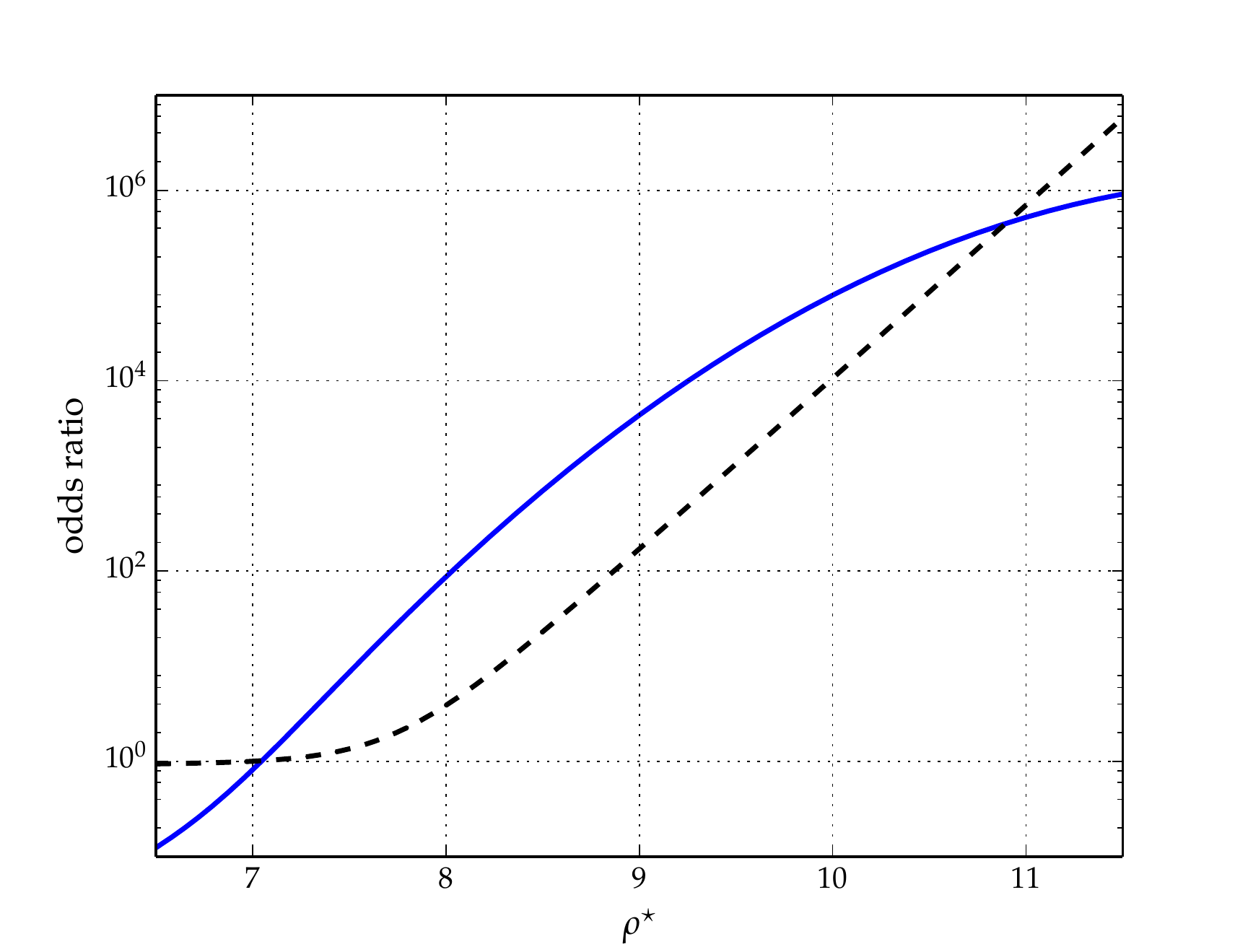}
\caption{\label{fig:odds}
The odds ratio between the signal and noise models.  
We consider two signal models: a GRB at unknown distance (black, dashed line)
and a GRB at a known distance (blue, solid line).  In this example, we have used the 
parameters from GRB080905A with a distance of 550 Mpc which gives an
expected SNR of 7.7. The blue curve gives the odds ratio for a BNS signal at that 
distance, as a function of SNR.  The black curve gives the odds ratio for a
BNS signal at an unknown distance.  At low SNR, knowledge of the distance increases
the odds ratio significantly; at higher SNRs it decreases the odds as the observed SNR is
no longer consistent with the distance.}
\end{center}
\end{figure}

The advanced detector network will, on average, be sensitive to a BNS merger
associated with a GRB within a distance of 400 Mpc, or $z \lesssim 0.1$.  The closest observed short
GRB is GRB 080905A with a measured redshift of $z=0.12$ \citep{Rowlinson:2010hf}.  
It is interesting to ask whether this GRB could have been observed by the 
advanced LIGO-Virgo network.  While the GRB was at a distance of 550 Mpc,
it was at a favourable sky location for the advanced LIGO and Virgo network and a BNS merger associated with this GRB 
may have been marginally detectable, but only once the redshift information is folded in.

Let us consider the expected distribution of the observed SNR in the gravitational wave search, 
under three distinct scenarios: 
no observed gravitational wave signal; a BNS merger signal associated with a short GRB at an unknown distance; 
a BNS merger signal at 550 Mpc.  
To obtain the distribution in the absence of a signal, we simply use the empirical estimate provided in Eq.~(\ref{eq:fap}).
For a signal at 550 Mpc in the direction of GRB080905A, a BNS merger will generate an expected network SNR of 7.7.  
The expected, maximum SNR observed in the gravitational wave search then follows a 
non-central $\chi^{2}$ with four degrees of freedom \citep{Harry:2010fr} overlaid on the noise background given in 
equation (\ref{eq:fap}). Finally, for a GRB with unmeasured redshift, we use the distance distribution as given in Eq.~\ref{eq:grb_prob},
i.e. signals distributed uniformly in $D^{3}$ at low redshift, with only a small probability of the GRB occurring within the 
LIGO-Virgo sensitive range.

In figure 4, we show the probability distribution for the SNR of the gravitational wave event under these three scenarios.  
The figure shows both the
probability distribution as well as the cumulative probability of observing an event above a given SNR.  In this
example, the knowledge of the distance greatly increases the chance of observing a signal with a moderate
SNR.  For example, the chance of observing an event with SNR $> 7.5$ due to 
noise alone is around 1\%, if there is a BNS merger at unknown distance it is 3\% whilst when the distance is known 
to be 550 Mpc it is 60\%.  

One way to visualise the benefit of a redshift measurement is through the odds ratio: 
the ratio of the signal probability to the noise probability.  This is plotted 
in Fig. \ref{fig:odds}.  For an observed SNR above 7.5, the signal model is favoured over the noise by a factor of 
10, increasing to 100 at SNR of 8.  Even at these low SNRs, this would be an interesting event.  However, 
if the distance is not known, a larger SNR (of 8.5 or 9) is required
before the signal model is strongly favoured over the noise.  Thus, if this GRB had occurred during the advanced 
detector era, there is a real chance that measuring the redshift would make the difference between identifying
a gravitational wave candidate and not.

\subsection{Constraining the jet opening angle}

A joint gravitational wave-GRB observation would provide a measurement of the binary's inclination angle and, consequently, 
would provide a constraint on the jet opening angle of GRBs.
However, the majority of observed gravitational wave signals are likely 
to be weak, with an SNR of ten or less, and this will make accurate parameter recovery difficult.
Accurate measurement of the binary inclination angle is further complicated by the fact that it is highly 
degenerate with the distance, particularly when the signal is close to face on.  Specifically, the
overall amplitude of the two polarisations scale as $\frac{(1 + \cos^{2} \iota)}{2D}$ and $\frac{| \cos \iota |}{D}$
and, at SNR of 10, we would expect to measure these amplitudes with roughly a 10\% accuracy.

For a face on signal (with $\iota \approx 0 \, \mathrm{or} \, \pi$), the two amplitudes are equal.  They differ
by 1\% for an inclination angle of $30^{\circ}$ and by 10\% for an inclination of $50^{\circ}$.  Thus, while the 
gravitational wave observation will constrain opening angles, it is most likely to limit the angle to be $\lesssim 45^{\circ}$.  
In the case where the redshift, and hence distance $D$, is known there will still be a $\sim 10$\% 
uncertainty in $\cos \iota$ corresponding to a constraint on the opening angle of $\lesssim 25^{\circ}$.
Even for the loudest signals, we are faced with an uncertainty in the Hubble constant of 1\% and a likely 
instrumental calibration error of 1\% or more \citep{Abadie2010223} making it difficult to constrain the opening angle 
to less than $10^{\circ}$.

\begin{deluxetable*}{rrrrrr}
\tablecolumns{6}
\tablecaption{The expected bounds on GRB opening angle during the early advanced LIGO-Virgo observing runs.  
These results assume that no gravitational wave signal is observed and use the observed GRB rate to 
infer the minimum jet opening angle consistent with the lack of gravitational wave detection.  The results in the two columns assume 
that all GRBs are either BNS (first column) or NSBH (2nd).  In both cases, the range quoted takes into account
both the uncertainty in the detector performance in these runs as well as the uncertainty in the local rate of GRBs.
\label{tab:opening}}
\tablehead{
\colhead{Epoch} & \colhead{Run Duration} & \multicolumn{2}{c}{BNS Range (Mpc)} & 
\multicolumn{2}{c}{limit on GRB opening angle (${}^{\circ}$)} \\
\cline{3-4} \cline{5-6} \\
\colhead{} &\colhead{} & \colhead{LIGO} & \colhead{Virgo} & \colhead{BNS} & \colhead{NSBH} }
\startdata 
2015        & 3 months & 40 - 80   &     -         & 0 - 3  & 0 - 6     \\
2016--17 & 6 months & 80 - 120 &  20 - 60  & 1  - 8  & 3 - 15   \\ 
2017--18 & 9 months & 120-170 &  60 - 85  & 3 - 15 & 7 - 35 
\enddata
\end{deluxetable*}

\begin{figure}[tbp]
\begin{center}
\includegraphics[width=\linewidth]{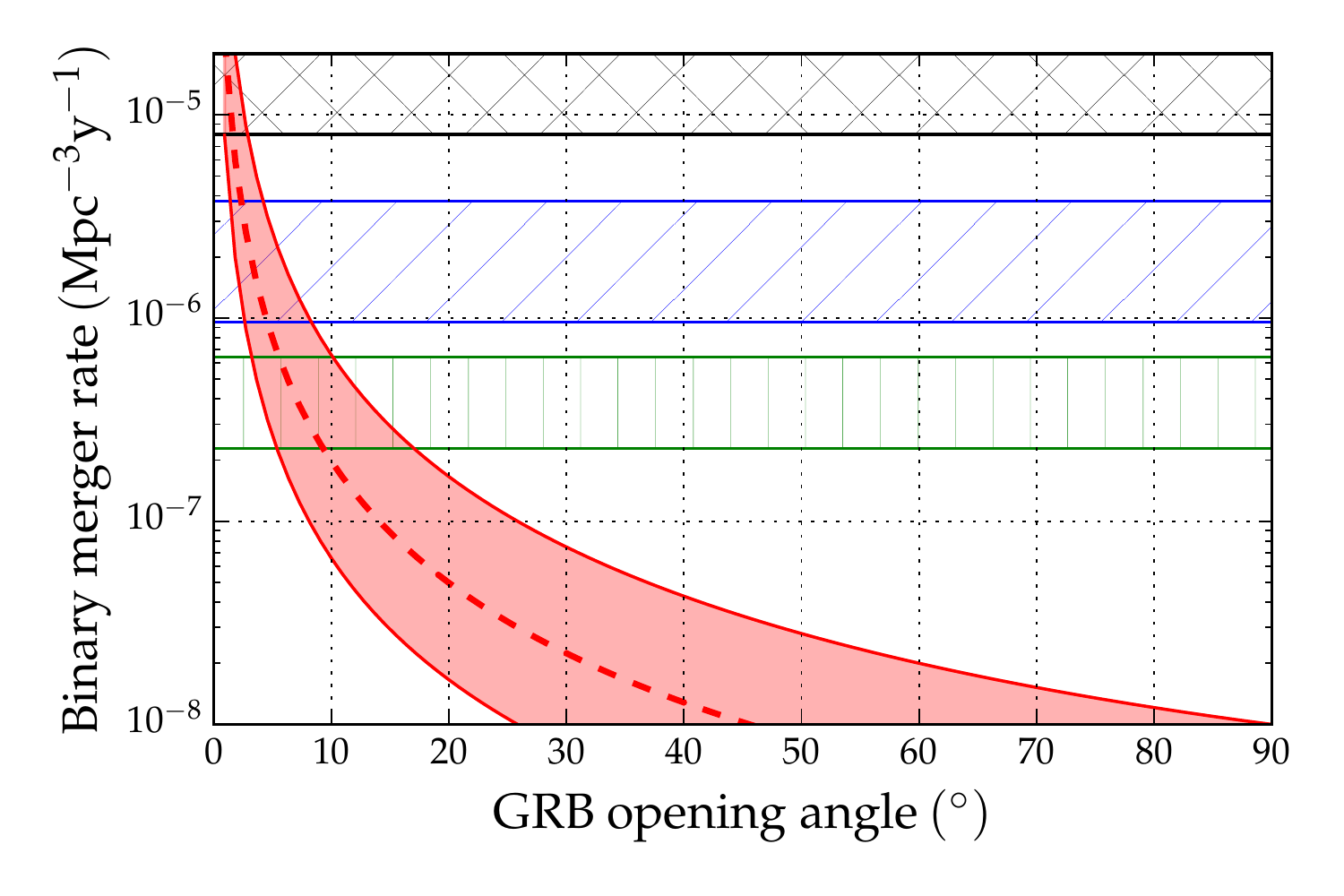}
\caption{\label{fig:bns_opening}
The expected upper limit on BNS rate for the planned observing runs, assuming that no GW events are 
observed: 2015 in black; 2016--17 in blue and 2017-18 in green. The ranges correspond to the uncertainties
in detector sensitivity as detailed in Table \ref{tab:nsns_rate}. 
The expected rate of GRB progenitors as a function of the GRB opening angle is overlaid in red.  
Assuming that all GRBs correspond to BNS, we can read off the lower limit on opening 
angle that would be obtained at the end of each run. }
\end{center}
\end{figure}

It is more likely that the observed populations of short GRBs and binary mergers will allow us to restrict
the opening angle of GRB jets.  It is clear from Figure \ref{fig:rate_prediction} how this would be done: if
the permitted range of BNS merger rates can be reduced from three orders of magnitude to a factor of 
two, then the GRB opening angle will be highly constrained.  First, consider the case where the
early science runs do not yield a gravitational wave detection.  We assume that, in the absence of a detection, 
the loudest event is consistent
with background and estimate the expected rate upper limit as $R^{\mathrm{ul}} = 2.3/VT$
 --- where $V$ is the volume searched and $T$ the time.
\citep{2008CQGra..25j5002B}.  From this, we can
read off the smallest opening angle consistent with the rate upper limit by re-arranging
equation (\ref{eq:rates}) to obtain
\begin{equation}
  1 - \cos\theta \ge \frac{R_{\mathrm{GRB}}}{f_{\gamma} R^{\mathrm{ul}}_{\mathrm{BNS}}} \, .
\end{equation}
Thus the tightest limit on $\theta$ is given by assuming the maximum BNS rate, i.e. right at the upper limit, 
and an $f_{\gamma}$ of unity, i.e that all BNS mergers produce GRBs.  
In Fig. \ref{fig:bns_opening} we plot
the expected upper limits, in the absence of a gravitational wave detection, from the early observing runs.  The 
bands here correspond to the uncertainties in detector sensitivities as given in 
Fig. \ref{tab:nsns_rate}.  For example, at the end of the 2016-17 run, the lack of a detection can place a constraint 
on the GRB opening angle between $2^{\circ}$ and $8^{\circ}$ depending upon the detector sensitivity achieved 
and the assumed GRB rate.  Thus, even in the absence of an observation, we are starting to impact  
measurements from GRB observations.
In Table \ref{tab:opening}, we summarise the results for BNS signals, and also 
for NSBH.  In both cases, we are assuming that all GRBs are produced by one particular type of merger.
This, of course, is unrealistic.  Although we cannot know the fraction of GRBs which have a BNS or
NSBH progenitor, we might reasonably assume that \textit{all} GRB 
progenitors are mergers.  Since the sensitivity to BNS mergers is less than NSBH, the conservative limit
comes from assuming that all mergers are BNS.  Alternatively, it is possible to make reasonable 
assumptions of priors for the various parameters and then marginalise over them to obtain a distribution 
for the opening angle \citep{Clark:2014}.

Of course, we hope to observe gravitational waves from binary mergers. Even a handful of observations will 
provide a measurement of the rate within a factor of two, which  will correspond to a much tighter horizontal 
band on figure \ref{fig:rate_prediction}.  
If, for example, the rate is $10^{-6} \mathrm{Mpc}^{-3} \mathrm{y}^{-1}$ then this will restrict the GRB 
opening angle to be between about $3$ and $8^{\circ}$.

\section{Conclusion}
\label{sec:discussion}

We have presented in detail the expectations for gravitational wave observations associated with 
short gamma ray bursts in the coming years.  The evidence for a binary merger progenitor of short GRBs 
is strong, and we have focused on this scenario.  By making use of the known time and sky location 
of the source, we have argued that it is appropriate to lower the gravitational wave search threshold by around 25\%
relative to the all sky, all time search.  This decrease in threshold will double the number of gravitational wave
events that can be detected in association with a GRB.  We have also demonstrated that if the redshift is measured 
(and is within the sensitive range of gravitational wave network), the detection threshold can be further reduced.  
Using this threshold reduction and the expected evolution of gravitational wave 
detector sensitivities given in \cite{Aasi:2013wya}, a joint gravitational wave--GRB 
observation is possible in the 2015 and 2016-17 observing
runs, but unlikely.  However, as the detectors approach 
their design sensitivity the rate of joint observations
increases and could be one or two per year for a BNS progenitor and even higher if the majority of GRBs have
NSBH progenitor. This, of course, depends critically upon the continued
operation of wide-field of view GRB satellites, such as Fermi, as well as the continued operation of the
InterPlanetary Network.  

The joint observation of gravitational wave and GRB signals will be a major milestone in 
understanding short GRBs, 
and will finally prove (or disprove) the binary merger progenitor scenario.  It will also shed light on the
GRB central engine by probing delays between the signals and a bound on the jet opening angle.
We have argued that the measurement of binary inclination from gravitational wave observations will have large 
uncertainties, due in part to the degeneracy with the measurement of distance, and is unlikely to constrain the opening angle
tightly.  However, an accurate measurement of the populations of both short GRBs and binary neutron star
(or neutron star-black hole) mergers will allow us to constrain the opening angle.  Even in the early advanced 
detector runs, we will be able to place lower bounds on the beaming angle of
short GRBs that will confront current observations.

\acknowledgements

The authors would like to thank the participants of the Royal Society International Scientific Seminar on Gravitational Waves and Gamma Ray Bursts for useful discussions.  In addition, they thank Ray Frey and Valeriu Predoi for comments on the manuscript.
This work was supported by the UK Science and Technologies Funding Council through grant ST/L000962/1, 
the Royal Society and the US National 
Science foundation through grants PHY-0847611, PHY-1205835, PHY-0955773 and AST-1333142.
                                                                                                                                             
\bibliographystyle{apj}
\bibliography{../references/refs}

\end{document}